\documentstyle[11pt,aaspp4]{article}
\begin{document}
\title{Implications of the radio afterglow from the $\gamma$-ray burst of 
May 8, 1997}
\author{Eli Waxman$^1$, Shri R. Kulkarni$^2$, Dale A. Frail$^3$}
\affil{$^1$Institute for Advanced Study, Princeton, NJ 08540}
\affil{$^2$Division of Physics, Mathematics an Astronomy 105-24, Caltech,
Pasadena CA 91125}
\affil{$^3$National Radio Astronomy Observatory, Socorro NM 87801}

\begin{abstract}

Radio observations of the afterglow of the $\gamma$-ray burst GRB970508 
provide unique new constraints on afterglow models. The quenching of 
diffractive scintillation at $\sim4$week delay provides the first direct 
estimate of source size and expansion rate. It implies an apparent size 
$R\sim10^{17}$cm and expansion at a speed comparable to that of light at 
$t\sim4$weeks, in agreement with the fireball model prediction 
$R=10^{17}(t/{\rm week})^{5/8}$cm. The radio flux and its dependence on time 
and frequency at 1--5 week delay are in agreement with the model and imply a 
fireball energy (assuming spherical symmetry) $\sim10^{52}$erg, consistent 
with the value inferred from observations at shorter delay. The observed radio 
behavior deviates from model predictions at delays $>5$weeks. This is expected,
since at this delay the fireball is in transition from highly-relativistic 
to sub-relativistic expansion, with Lorentz factor $\gamma\le2$. Deviation may 
be due to a change in the physical processes associated with the shock wave as 
it becomes sub-relativistic (e.g. a decrease in the fraction of energy carried
by magnetic field), or to the fireball being a cone of opening angle 
$\sim1/\gamma\sim1/2$. We predict the future behavior of the radio flux 
assuming that the latter interpretation is valid. These predictions may be 
tested by radio observations in the frequency range 0.1--10GHz on time scale 
of months.

\end{abstract}

\section{Introduction}

The availability of accurate positions for GRBs from the BeppoSAX satellite
(\cite{bs1,bs2a,bs2b,bs3}) 
allowed for the first time to detect delayed emission
associated with GRBs in X-ray (\cite{bs1},b,\cite{bs2a,bepx1},b), 
optical (\cite{opt1,opt1a,Paradijs,O6654,Onat}) 
and radio (\cite{R}) wave-bands. The detection of absorption lines in the
optical afterglow of GRB970508 provided the first direct estimate of
source distance, constraining the redshift of GRB970508 to $0.86<z<2.3$ 
(\cite{z}). Observed X-ray to radio afterglow are most naturally explained
by models based on relativistic blast-waves at cosmological distances
(\cite{PRAG,MRAG,VAG,AGW},b,\cite{AGWMR}). Using these models,
combined radio and optical data allowed for the first time 
to directly estimate the total GRB energy, implying an energy of 
$\sim10^{52}{\rm erg}$ (assuming spherical symmetry) for GRB970508 
(\cite{AGWb}).

Radio observations of the afterglow of GRB970508 (\cite{VLA,VLBI}) provide
unique new information on the afterglow source. Shortly after the first
detection of GRB radio afterglow, it was pointed out by Goodman 
(\cite{Goodman}) that
if the source angular size is as small as predicted by fireball models, 
$\sim1\mu{\rm as}$, then 
the radio flux should be modulated by scintillation due to the local 
inter-stellar medium. The predicted modulation has been observed
(\cite{VLA,VLBI}) and it provides for the first time direct constraints
on source size and expansion rate. We discuss in \S3 the implications 
to fireball models of the observed modulation. 
In \S4 we discuss the implications of the
information on long term afterglow behavior provided by the 
radio monitoring of GRB970508 over three months. Deviation from
model predictions is expected on this time scale (\cite{AGWb}), 
as the fireball decelerates from highly-relativistic to marginally 
relativistic speed. However, the lack of a theory describing the relevant 
physical processes associated with the shock wave (magnetic field generation, 
energy transfer to electrons) does not allow a unique interpretation of the 
observed behavior. We derive equations describing the fireball evolution 
in the non-relativistic regime, which allow to demonstrate that the 
observed behavior can not be accounted for by the deviation of the
hydrodynamic evolution from that predicted by the highly relativistic 
scalings. We discuss possible interpretations of the long term
behavior, and predict the future fireball radio emission
under the assumption that the observed behavior is due to a finite 
opening angle of the fireball. We summarize our conclusions and predictions in
\S5. 

We note here that two types of 
fireball models were considered in the literature as possible interpretations
of existing GRB970508 afterglow data. In the ``adiabatic'' model (\cite{AGWb})
the fireball radiates over time only a small fraction of its energy, which is 
of order $10^{52}$erg. 
In the ``radiative'' models (\cite{V2,KP}) the fireball radiates most of its 
energy over a short time, and its energy is reduced to 
$\sim10^{49}$erg on time scale of  days. 
We discuss in this paper mainly implications of radio observations to the 
adiabatic model since, as we demonstrate in \S4, long term
radio observations rule out the radiative models, and since,
as we briefly show in \S2 below, the radiative models are not
self-consistent.

\section{Radiative versus non-radiative models}

In fireball afterglow models, a highly relativistic shell
encounters, after producing the GRB, some external medium. As the shell
decelerates it drives a relativistic shock into the surrounding medium. This
shock continuously heats fresh gas and accelerates relativistic electrons,
which produce the observed radiation through synchrotron emission.
The magnetic field $B$ behind the shock and the characteristic Lorentz
factor $\gamma_e$ of the electrons are determined in these models by assuming 
that fractions $\xi_e$ and $\xi_B$ 
of the fireball energy are carried by electrons and magnetic field 
respectively. This implies $B^2/8\pi=4\xi_B \gamma^2 n m_p c^2$ and 
$\gamma_e=\xi_e\gamma m_p/m_e$ (here $n$ is the number density of the
surrounding medium, and $\gamma$ is the fireball Lorentz factor). 
There is no theory which allows to determine the values of $\xi_e$ and
$\xi_B$. Afterglow observations are consistent with $\xi_e\sim\xi_B\sim0.1$ 
during the relativistic expansion of the fireball (\cite{AGWb}).

The underlying assumption of radiative models is that all 
the kinetic energy lost by the fireball as it decelerates is radiated away. 
This requires an efficient process converting the kinetic energy flux
to electron thermal energy, and requires the electron cooling time to be much 
shorter than the dynamical time, i.e. the time scale for expansion in
the fireball rest frame
$\tau_d=r/4\gamma c$ (here $r$ is the fireball radius). 
The ratio of synchrotron cooling time,
$\tau_s=6\pi m_e c/\sigma_T\gamma_e B^2$, to dynamical time is
\begin{equation}
\Theta\equiv{\tau_s\over\tau_d}=
{3\over4}\left({m_e\over m_p}\right)^2(\xi_e \xi_B\sigma_T n r\gamma^2)^{-1}=
16(\xi_e\xi_B/0.25)^{-1}(n_1 r_{17} \gamma^2)^{-1},
\label{Theta}
\end{equation}
where $n=1n_1{\rm cm}^{-3}$, $r=10^{17}r_{17}$cm. (This result is similar to
that given in [\cite{V2}]). In the radiative model
of Katz \& Piran (\cite{KP}) the fireball expands into a uniform
density medium, $n_1=1$, and due to rapid energy loss the fireball becomes
non relativistic at $6$ day delay with $r_{17}\sim\gamma\sim1$.
In the radiative model of Vietri (\cite{V2}), the ambient medium density
drops with radius as $r^{-2}$ allowing the fireball to remain relativistic 
as it loses energy. In this case $n_1\sim10^{-3}$ and 
$r_{17}\sim\gamma\sim3$ at $t=6$day. 
In both models, therefore, $\Theta\gg1$ at this
time. Furthermore, the time dependence of $\Theta$, $\Theta\propto t^{1}$
for $n\propto r^{-2}$ and $\Theta\propto t^{5/7}$ for uniform density,
implies that the assumption $\Theta\ll1$ is not valid in both models for
$t\gtrsim1$hr.

\section{Source size and expansion rate}

Due to relativistic beaming, the radiation from a relativistic fireball
seen by a distant observer is emitted from a cone of the fireball around the 
source-observer line of sight, with an opening angle $\sim1/\gamma$. The 
apparent radius of the emitting cone is $R=r/\gamma$, where $r$ is the 
fireball radius, and photons emitted from such a cone are delayed, 
compared to those emitted on the line of sight, by $t=r/2\gamma^2c$.
Thus, the apparent radius of the fireball
is $R=2\gamma(r) ct$ (a detailed calculation of fireball emission [\cite{W16}]
introduces only a small correction, $R=1.9\gamma(r) ct$), where $r$ and $t$ 
are related by $t=r/2\gamma^2c$.
Using eqs. (1) \& (2) of (\cite{AGWb}) we have
\begin{equation}
\gamma=4\left({1+z\over2}\right)^{3/8}\left({E_{52}\over n_1}
\right)^{1/8}t^{-3/8}_{\rm w},
\label{rg}
\end{equation}
and
\begin{equation}
R=8\times10^{16}\left({1+z\over2}\right)^{5/8}\left({E_{52}\over n_1}
\right)^{1/8}t^{5/8}_{\rm w}{\rm cm}.
\label{R}
\end{equation}
Here $E=10^{52}E_{52}$erg is the fireball energy and $t=1t_{\rm w}$week.
(Note, that the relation
$t=r/2\gamma^2c$ should not be replaced by
$t=r/16\gamma^2c$, as recently argued in [\cite{Sari}], 
since the latter relation holds only for the arrival
time of photons which are emitted on the line of sight and not valid for most 
of the photons, which are emitted from a cone of opening angle $\sim1/\gamma$ 
[\cite{W16}]).

Scattering by irregularities in the local interstellar medium (ISM) may 
modulate the observed fireball radio flux (\cite{Goodman}). If scattering
produces multiple images of the source, interference between the multiple
images may produce a diffraction pattern (on an imaginary plane perpendicular
to the line of sight), leading to strong variations of the flux as the observer
moves through the pattern. Two conditions need to be met in order
to produce such diffractive scintillation: (i) The scattering should be strong
enough to produce multiple images; (ii) The source size should be small enough
so that different points on the source produce similar diffraction patterns.
There is significant observational evidence that ISM electron density
fluctuations are described by a power-law spectrum, 
$\langle\delta N_e(\vec k)\delta N_e(-\vec k)\rangle=C_N^2 k^{-11/3}$, where
$\vec k$ is the spatial wave number. For this distribution, the characteristic
deflection angle is given by
$\theta_d=2.34\lambda^{11/5}r_e^{6/5}(SM)^{3/5}$, where $\lambda$ is the
wavelength, $r_e$ the classical electron radius, and the scattering
measure $SM$ is the integral of $C_N^2$ along the line of sight.
In the frame work of 
the ``thin screen'' approximations, i.e. assuming that all scattering occurs 
in a narrow layer at distance $d_{\rm sc}$, multiple images are produced
for frequencies (\cite{Goodman})
\begin{equation}
\nu\le11d_{\rm sc,\,kpc}^{6/17}
\left({SM\over10^{-3.5}{\rm m}^{-20/3}{\rm kpc}}\right)^{5/17}{\rm GHz}.
\label{nuss}
\end{equation}
Here, values were chosen for the ISM scattering properties $SM$ and 
$d_{\rm sc}$, which are typical for sources at high Galactic latitude 
($b=27\arcdeg$ for GRB970508). The characteristic length scale of the 
diffraction pattern is
\begin{equation}
dx={\lambda\over2\pi\theta_d}=
3.3\times10^{10}\nu_{10}^{6/5}
\left({SM\over10^{-3.5}{\rm m}^{-20/3}{\rm kpc}}\right)^{-3/5}{\rm cm}.
\label{dx}
\end{equation}
In order for the diffraction patterns produced by different points on the 
source to be similar, so that the pattern is not smoothed out due
to large source size, the angular source size $\theta_s$ should satisfy 
$\theta_s d_{\rm sc}<dx$. For a source at redshift
$z=1$ this requirement implies an upper limit to the apparent source size
\begin{equation}
R<1.0\times10^{17}{\nu_{10}^{6/5}\over d_{\rm sc,\,kpc}h_{75}}
\left({SM\over10^{-3.5}{\rm m}^{-20/3}{\rm kpc}}\right)^{-3/5}\,{\rm cm},
\label{Rdiff}
\end{equation}
where $\nu=10\nu_{10}$GHz, $h_{75}$ is the Hubble constant in units of
$75{\rm km/s\, Mpc}$. Due to the weak dependence of the angular diameter 
distance on $z$, the upper limit on $R$ is not sensitive to source redshift.
The main uncertainty in (\ref{Rdiff}) is due to uncertainty in the 
scattering properties of the ISM. 
Combining (\ref{Rdiff}) and (\ref{nuss}) we find that 
\begin{equation}
R<1.1\times10^{17}d_{\rm sc,\,kpc}^{-11/17}h^{-1}_{75}
\left({SM\over10^{-3.5}{\rm m}^{-20/3}{\rm kpc}}\right)^{-3/17}\,{\rm cm}
\label{Rmax}
\end{equation}
is required to allow diffractive scintillation. 
Due to variations in ISM scattering
properties along different lines of sight, the numerical
value in (\ref{Rmax}) is accurate to a factor of a few.

The frequency range $\Delta\nu$
over which the diffraction pattern is similar, and therefore over which
flux modulation is correlated, is
\begin{equation}
\Delta\nu={c\over2\pi\theta_d^2d_{\rm sc}}=
0.4d_{\rm sc,\,kpc}^{-1}
\left({SM\over10^{-3.5}{\rm m}^{-20/3}{\rm kpc}}\right)^{-6/5}
\left({\nu\over5{\rm GHz}}\right)^{22/5}{\rm GHz}.
\label{dnu}
\end{equation}
For a characteristic velocity through the diffraction pattern of 
$\simeq 30{\rm km\,s}^{-1}$, due to Earth's
orbital motion and to the Sun's peculiar velocity, (\ref{dx}) implies a
time scale for variations $\simeq3$hr.

Comparing (\ref{R}) and (\ref{Rmax}) we find that on time scale of weeks
the apparent fireball size is comparable to the maximum size for which
diffractive scintillation is possible. On shorter time scale, therefore,
strong modulation of the radio flux is expected. On longer time scale
we expect diffractive scintillation to be quenched due to large source size.
When diffractive scintillation is quenched, the flux is nevertheless expected
to be modulated due to refraction (i.e. due to focusing/defocusing of rays).
However, the modulation amplitude should decrease (to $\sim10\%$, 
\cite{Goodman}), and modulation should be correlated over a wide frequency
range. 
Figure 1 presents the light curves of the radio afterglow at 8.46GHz, 4.86GHz
and 1.43GHz. Figure 2 presents the fluxes 
at 4.86GHz as a function of the flux at
8.46GHz. Strong modulation of the radio flux is observed during the first
month, accompanied by strong variations in the ratio of flux at 8.46GHz and
4.86GHz. This behavior
is consistent with that expected due to diffractive scintillation. The 
radio flux is not sampled at a sufficient rate to determine whether the
variability time scale is consistent with that expected for diffractive 
scintillation. At delays longer than $\sim1$ month the modulation 
amplitude decreases, and the flux ratio is consistent with being constant.
This is consistent with quenching of diffractive scintillation due to increased
source size.

Observations are therefore in agreement with fireball model predictions.
They imply that the source size is close to the upper limit given by
(\ref{Rmax}) after $\sim1$month. This is consistent with (\ref{R}) and
implies expansion at a speed comparable to that of light. Due to the very weak
dependence of $R$ on fireball model parameters it is
not possible to accurately determine parameters based on the quenching of  
diffractive scintillation. On the other-hand, since 
$R$ is very insensitive to model parameters, reducing the uncertainty in the 
size estimate based on scintillation, by reducing the 
uncertainty in ISM scattering properties towards the GRB, would
provide a stringent test of the fireball model.

\section{Long term behavior}

\subsection{Relativistic regime}

The radio light curves are compared with the predictions of the adiabatic
fireball model (\cite{AGWb}) in Figure 1. In this model, the observed
frequency at which the synchrotron spectral intensity peaks is
\begin{equation}
\nu_m^R=5\times10^{12}\left({1+z\over2}\right)^{1/2}
(\xi_e/0.2)^2(\xi_B/0.1)^{1/2}E_{52}^{1/2}t_{\rm w}^{-3/2}{\rm Hz},
\label{num}
\end{equation}
and the observed intensity at $\nu_m$ is
\begin{equation}
F_{\nu_m}^R=1\,\left({1 + z\over 2}\right)^{-1}\left[{1 - 1/\sqrt{2}\over
1 - 1/\sqrt{1 + z}}\right]^{2}
n_1^{1/2}(\xi_B/0.1)^{1/2}E_{52} {\rm mJy}.
\label{Fm}
\end{equation}
The superscript $R$ implies that the expressions are valid for the
highly relativistic regime.
The flux at $\nu_m$ is produced by electrons at the characteristic electron
energy, $\varepsilon=\gamma_e m_e c^2$. 
The flux at higher frequency is produced by 
higher energy electrons. For a power-law electron spectrum, 
$dN_e/d\varepsilon_e\propto\varepsilon^{-p}$, $F_\nu\propto\nu^{-(p-1)/2}$
at $\nu>\nu_m$. Typical parameters required to fit observations are
$E_{52}\sim n_1\sim1$, $\xi_e\sim\xi_B\sim0.1$, and $p\sim2$.

The flux at low frequency, $\nu<\nu_m$, is due to the extension of synchrotron
emission of electrons at $\varepsilon=\gamma_e m_e c^2$ to frequencies
$\nu<\nu_m$ [$F_\nu\propto(\nu/\nu_m)^{1/3}$ at $\nu\ll\nu_m$].
At low frequency self-absorption becomes significant.
The self-absorption frequency, where the fireball optical depth is unity, is
\begin{equation}
\nu_A^R=1\left({1+z\over2}\right)^{-1}(\xi_e/0.2)^{-1}(\xi_B/0.1)^{1/5}
E_{52}^{1/5}n_1^{3/5}{\rm\, GHz},
\label{nuA}
\end{equation}
and at $\nu<\nu_m$ the optical depth is given by $\tau_\nu=(\nu/\nu_A)^{-5/3}$.
Self-absorption reduces the flux by a factor $(1-e^{-\tau_\nu})/\tau_\nu$.

The solid smooth curves in Figure 1 give the model fluxes for $E_{52}=n_1=2$,
$\xi_e=0.2$ and $\xi_B=0.1$. These are essentially the same parameter values 
inferred in (\cite{AGWb}) from optical and radio afterglow data at
delays $t\le6$day. Although many simplifying assumptions were made, in order
to obtain a simple description of fireball behavior, the model is in agreement 
with the data obtained during the first $\sim5$ weeks.
At later time model curves deviate from the data. This behavior is expected 
(\cite{AGWb}), as the fireball decelerates from highly-relativistic to 
marginally relativistic speed. At this stage the scalings (\ref{rg},\ref{R}) 
do not
give an accurate description of the fireball dynamics. Furthermore, there 
is no theory which allows to determine the parameters $\xi_e$ and $\xi_B$.
These parameters may change as the shock decelerates, thus affecting the 
predictions (\ref{num}--\ref{nuA}). Prior to discussing in \S5 the possible 
implications of observations at $t>5$weeks we derive in \S4.2 the 
equations describing the fireball dynamics at the non-relativistic stage.
This would allow to estimate the effects due to deviation from the 
scaling laws (\ref{rg},\ref{R}).

\subsection{Transition to the non-relativistic regime}

As the fireball becomes non-relativistic its expansion approaches that
described by the Sedov-von Neumann-Taylor solutions (\cite{Sedov,vN,Taylor}).
At this stage the shock radius is given by 
$r=\xi_0(\hat\gamma)(Et^2/nm_pc^2)^{1/5}$, where $\xi_0$ is a function of the
adiabatic index of the gas $\hat\gamma$. $\xi_0=0.99$ for 
$\hat\gamma=4/3$ (relativistic fluid) and $\xi_0=1.15$ for 
$\hat\gamma=5/3$ (non-relativistic limit). The non-relativistic behavior may
be described as
\begin{equation}
\beta\equiv\dot r/c=\xi_0^{5/2}(r/r_{NR})^{-3/2}, \quad 
r/r_{NR}=(\xi_0/1.15)(t/t_{NR})^{2/5},
\label{rb}
\end{equation}
by defining 
\begin{equation}
r_{NR}=1.0\times10^{18}\left({E_{52}\over n_1}\right)^{1/3} {\rm cm}, \quad
t_{NR}=24{1+z\over2}\left({E_{52}\over n_1}\right)^{1/3} {\rm week}.
\label{NR}
\end{equation}
With these definitions, the time dependence (\ref{rg}) of the Lorentz
factor in the highly-relativistic regime may be written as 
$\gamma=1.3(t/t_{NR})^{-3/8}$. Thus, the relativistic solution is valid for
$t\ll t_{NR}$, where $\gamma^2\gg1$, and the non-relativistic solution for 
$t\gg t_{NR}$, where $\beta^2\ll1$

For the non-relativistic regime, assuming that fractions $\xi_B$
and $\xi_e$ of the dissipated energy are carried by magnetic field
and electrons imply $B^2/8\pi=\beta^2nm_pc^2$ and 
characteristic electron Lorentz factor
$\gamma_e=(m_p/2m_e)\xi_e\beta^2$. The frequency at which the synchrotron
intensity peaks, $\nu_m=\gamma_e^2eB/2\pi m_e c$, is
\begin{equation}
\nu_m^{NR}=4\left({1+z\over2}\right)^{-1}
(\xi_e/0.2)^2(\xi_B/0.1)^{1/2}n_1\beta^5{\rm GHz}.
\label{num_nr}
\end{equation}
The intensity at $\nu_m$ is $F_{\nu_m}^{NR}=
N_e (\sqrt{3}e^3 B/2\pi m_e c^2) (1+z)/4\pi d_L^2$, 
where $N_e=4\pi r^3n/3$ is the number
of radiating electrons, $d_L$ the luminosity distance. Using (\ref{Fm})
we have
\begin{equation}
F_{\nu_m}^{NR}=F_{\nu_m}^{R}\beta\left({r\over r_{NR}}\right)^3.
\label{Fm_nr}
\end{equation}
Extrapolation of the non-relativistic expression (\ref{Fm_nr}) to $t=t_{NR}$
gives a peak flux similar to that given by (\ref{Fm}) for the relativistic 
regime. This is expected, since in the relativistic regime $F_{\nu_m}$ is
independent of time, and therefore of $\gamma$. Thus,
as the fireball decelerates to non-relativistic speed the peak
flux is approximately given by the relativistic expression (\ref{Fm}).
At later time, $t\gg t_{NR}$, $F_{\nu_m}\propto(t/t_{NR})^{3/5}$. 
Extrapolation of the relativistic expression (\ref{num}) for $\nu_m$ to 
$t=t_{NR}$ gives $\nu_m=100(\xi_e/0.2)^2(\xi_B/0.1)^{1/2}n_1^{1/2}$GHz,
significantly higher than the extrapolation of the non-relativistic
expression (\ref{num_nr}). Thus, as the fireball decelerates $\nu_m$
decreases with time faster than given by (\ref{num}), and 
$\nu_m\propto(t/t_{NR})^{-3}$ for $t\gg t_{NR}$.

The self-absorption frequency, where the fireball optical depth is unity, is
\begin{equation}
\nu_A^{NR}=\nu_A^{R}\beta^{-8/5}\left({r\over r_{NR}}\right)^{3/5}.
\label{nuA_nr}
\end{equation}
Comparing (\ref{nuA_nr}) and (\ref{nuA}) we find that the self-absorption
frequency is time independent and is given by the relativistic expression
for $t<t_{NR}$. At later time (\ref{nuA_nr}) implies that the self-absorption
frequency increases. However, (\ref{nuA_nr}) is valid only as long as
$\nu_A<\nu_m$. Since the dependence of $\nu_m$ on time for $t>t_{NR}$ 
is stronger than that of $\nu_A$, $\nu_m$ drops below $\nu_A$ when $\nu_A$
is not significantly higher than the value given by (\ref{nuA}). At later
time $\nu_A$ decreases with time. For an electron spectrum 
$dN_e/d\varepsilon_e\propto\varepsilon_e^{-2}$, 
$\nu_A\propto(t/t_{NR})^{-2/3}$.

\section{Implications}

\subsection{Ruling out radiative expansion}

From the analysis of \S4.2, the relativistic expression (\ref{Fm})
is a good approximation for the fireball peak flux not only for 
$\gamma\gg1$, but as long as $\beta\sim1$. Since radio observations imply
that the fireball expands with $\beta\sim1$ on time scale of weeks,
the observed flux of order 1mJy implies $E_{52}n^{1/2}_1\gtrsim1$ on weeks
times scale. This rules out the radiative models, where the fireball energy
decreases to $10^{49}$erg on day time scale. We note here that it was argued
in (\cite{KP}) that a mildly relativistic fireball, $\gamma-1\sim1$, with
$E\sim10^{49}$erg would produce the observed $\sim0.1$mJy flux at 1.43GHz
at $\sim6$day delay. This is in contradiction with our results, 
(\ref{Fm}) and (\ref{Fm_nr}), which imply a much lower flux. The (\cite{KP}) 
derivation is, however, not self consistent. The flux at 1.43GHz is obtained
in (\cite{KP})
using the Rayleigh-Jeans law, for which $F_\nu\propto\nu^2$ and which is 
valid only for high optical depth. From (\ref{nuA}) and (\ref{nuA_nr})
the self-absorption frequency for the parameters
chosen in (\cite{KP}), $E\sim10^{49}$erg and $n_1=1$,
is $\simeq0.2$GHz. Using the Rayleigh-Jeans formula for the flux
at 1.43GHz, where $\tau_\nu\simeq0.05$, overestimates the flux by a factor
$(1-e^{-\tau_\nu})^{-1}\sim30$.

\subsection{Deviations during the transition to non relativistic expansion}

On time scale $\gtrsim5$weeks the observed radio behavior deviates from model
predictions. The flux at 4.86GHz and 8.46GHz is significantly below model
predictions at $t\sim10$weeks. The increase of flux at 1.43GHz at this
time indicates that the frequency $\nu_m$ at which the intensity peaks
drops to $\sim5$GHz at $t\sim10$week, where (\ref{num}) predicts 
$\nu_m\sim100$GHz. 
From the analysis of the previous section, this behavior
can not be explained based only on the deviation from the highly-relativistic
scaling laws (\ref{rg},\ref{R}) 
as the fireball decelerates to mildly relativistic
velocity, $\gamma-1\sim1$. The synchrotron peak intensity is not expected to 
decrease [cf. eq. (\ref{Fm_nr})], and therefore the decrease in 4.86GHz and
8.46GHz flux can not be accounted for. 
The peak frequency $\nu_m$ is expected to
decrease faster than predicted by (\ref{num}). However, it is expected to
decrease to $\sim5$GHz only when the fireball becomes non-relativistic,
i.e. at $t\sim t_{NR}\sim24$week [cf. eqs. (\ref{num_nr}), (\ref{rb})].
Note, that a change in the ambient medium density $n$ is not likely to
account for the observed behavior. A decrease in flux may result from
a decrease in $n$ (\ref{Fm}). However, the peak frequency (\ref{num})
is independent of $n$, and the apparent decrease in $\nu_m$ can not be
accounted for.

Clearly, the observed behavior may be explained by deviations of the 
equipartition fractions $\xi_e$ and $\xi_B$ from the values 
$\xi_e\sim\xi_B\sim0.1$, which are implied by observations at $t<5$weeks. 
Due to the lack of a theory determining these parameters, it is not possible
to predict their dependence on shock Lorentz factor. Thus, if the observed
behavior is due to changes in $\xi_e$ and $\xi_B$, it is difficult to
predict the future fireball behavior.
However, the observed deviations from the model may also result from a 
different effect. This is discussed below.

\subsection{Non-spherical fireballs}

We have so far assumed that the fireball is spherically
symmetric. The results are valid also for the case where the fireball
is a cone of finite opening angle $\theta$, as long as $\gamma>1/\theta$.
In this case, the fireball energy $E$ in the equations should be understood
as the energy the fireball would have had if it were spherically symmetric.
The actual fireball energy is $E'\simeq\theta^2E/2$.
Deviations from the spherical model would appear at late time, as 
$\gamma$ decreases below $1/\theta$ and the fireball starts expanding 
transversely as well as radially. Let us briefly discuss the expected
behavior at later time. 

After a transition phase the fireball would
approach spherically symmetric expansion, which may again be described by
the equations derived above, with $E$ replaced by the actual fireball
energy $E'$. Numerical calculations would probably be 
required to describe the fireball evolution in the stage it approaches
spherically symmetric behavior. Qualitatively,
as the fireball expands transversely its energy per unit solid angle
along the line of sight decreases. Thus, it would appear as if the fireball 
energy is decreasing with time. This would lead to decrease in the peak flux
[cf. (\ref{Fm})] and in the peak frequency [cf. (\ref{num})], in 
qualitative agreement with the observed trends. The time scale for transition
to spherically symmetric behavior may be estimated as follows. 
The scaling laws (\ref{rg}) are derived
from energy conservation, $E\simeq\gamma^2(4\pi r^3/3)nm_pc^2$. This
implies that at the radius $r_\theta$, where $\gamma=1/\theta$, the rest
mass energy contained in the sphere with $r=r_\theta$ is comparable to
the total fireball energy $E'$. Thus, as the fireball approaches
spherically symmetric behavior it necessarily becomes sub-relativistic, and
at later time it is described by (\ref{num_nr}--\ref{nuA_nr}), with $E$
replaced by $E'$. The transition to spherical non-relativistic behavior 
occurs on a time scale [cf. eq. (\ref{NR})]
$t_{NR}=18(\theta^2E_{52}/n_1)^{1/3}$week.

The deviation at $t\sim5$week from model predictions may therefore be
accounted for by the fireball being a cone of opening angle
$\theta\sim1/\gamma(5{\rm week})\sim1/2$. This implies that the ``real''
fireball energy is $E'\sim2\times10^{51}$erg. In this case, the decrease in 
flux and in peak frequency are accounted for by the transition to 
spherical non-relativistic
behavior. The transition should take place on a time scale 
$t_{NR}\sim12$week. On this time scale the peak frequency is expected to
decrease to $\sim4$GHz [cf. eq. (\ref{num_nr})], and the peak intensity to 
$\sim0.3$mJy [cf. eq. (\ref{Fm_nr})]. This is in agreement with
the observed behavior. The fireball behavior for
$t\gg12$week is described by (\ref{rb}--\ref{nuA_nr}), with
$E_{52}\simeq0.2$, $n_1=2$.

\section{Conclusions}

Comparison of radio observations of the afterglow of the $\gamma$-ray burst 
GRB970508 with fireball model predictions lead to the following conclusions:
\begin{itemize}
\item{} The source size implied by the quenching of diffractive
scintillation at $\sim4$week, $\sim10^{17}$cm,
and the inferred expansion at a speed comparable to that of light
are consistent with published (\cite{AGWb}) model predictions (\ref{rg}),
(\ref{R}).
\item{} The radio 
flux and its dependence on time and frequency at 1--5 week delay are in
agreement with the model (Figure 1)
and imply a fireball energy (assuming spherical 
symmetry) $\sim10^{52}$erg. This is consistent with the value inferred from 
observations at shorter delay (\cite{AGWb}), and rules out ``radiative''
models (\cite{V2}, \cite{KP}), where fireball energy is reduced to 
$\sim10^{49}$erg on day time scale. 
\item{} The deviation of observed radio behavior from
model predictions at delays $>4$weeks is expected, as on this time scale 
the fireball is in transition from highly-relativistic 
to sub-relativistic expansion, with Lorentz factor $\gamma\le2$.
We have shown that the observed behavior can not be accounted for by 
the deviation of the hydrodynamic behavior from that predicted by the highly
relativistic scalings (\ref{rg},\ref{R}), 
or by changes in ambient medium density.
\item{} The observed behavior may be explained by deviation of the 
equipartition fractions $\xi_e$ and $\xi_B$ from the values 
$\xi_e\sim\xi_B\sim0.1$, which are implied by observations at $t<5$weeks. 
Due to the lack of a theory determining these parameters, it is not possible
to predict their dependence on shock Lorentz factor. Thus, if the observed
behavior is due to changes in $\xi_e$ and $\xi_B$, it is difficult to
predict the future fireball behavior.
\item{} However, the observed behavior at $t>5$week may also be accounted for
by the fireball being a cone of finite opening angle,
$\theta\sim1/\gamma(5{\rm week})\sim1/2$, which implies that the ``real''
fireball energy is $E'\simeq\theta^2E/2\sim2\times10^{51}$erg. 
In this case, at $t\sim5$week the fireball rapidly expands transversely,
leading to a decrease in the energy per solid angle along the line of sight.
This may account for the observed decrease in peak flux and peak frequency.
On time scale $t_{NR}\sim12$week [cf. eq. (\ref{NR})] the fireball
approaches spherical non-relativistic expansion.
On this time scale the peak frequency is expected to
decrease to $\sim4$GHz [cf. eq. (\ref{num_nr})], and the peak intensity to 
$\sim0.3$mJy [cf. eq. (\ref{Fm_nr})]. This is in agreement with
the observed behavior.
\end{itemize}

If the observed behavior is indeed due to the fireball being a cone of
finite opening angle, than the future  behavior may be predicted.
On time scale of $\sim12$week, the fireball should approach spherical
non-relativistic expansion. At times $t\gg12$week the behavior is given
by eqs. (\ref{rb}--\ref{nuA_nr}), which
describe the spherical non-relativistic behavior, with
$E_{52}\simeq0.2$, $n_1=2$. The frequency at which the intensity peaks should
decrease with time, $\nu_m\propto t^{-3}$, dropping to 0.3GHz in $\sim0.5$yr.
The peak intensity should be $\sim0.3$mJy over this period. If the electron
energy distribution is similar to the distribution inferred for the 
relativistic regime, $dN_e/d\varepsilon\propto\varepsilon^{-p}$ with
$p\sim2$, then the flux at $\nu>\nu_m$ should decrease approximately
as $t^{-1}$, similar to the highly relativistic case, and at a given
time the intensity should drop with frequency as $\nu^{-\alpha}$ with
$\alpha\sim0.5$. 

The time dependence of the flux at $\nu>\nu_m$ is similar in the 
non-relativistic and the relativistic regime. At early time, $t<1$week,
the optical flux decrease as $t^{-1}$ (\cite{Onat}). However, the optical 
flux should drop at $\sim5$week below the $t^{-1}$ extrapolation from
early time, as the effective fireball energy decreases. The
flux expected at $\sim10$week in this model is $m_R\sim25$. 
Agreement of the observed optical behavior with the behavior described above
would provide support to the model discussed here, where it is assumed that
both optical and radio fluxes are produced by the expanding fireball.

\paragraph*{Acknowledgments.} 

E. Waxman acknowledges support by a W. M. Keck Foundation grant 
and NSF grant PHY95-13835.

\newpage

\begin{figure}
\plotone{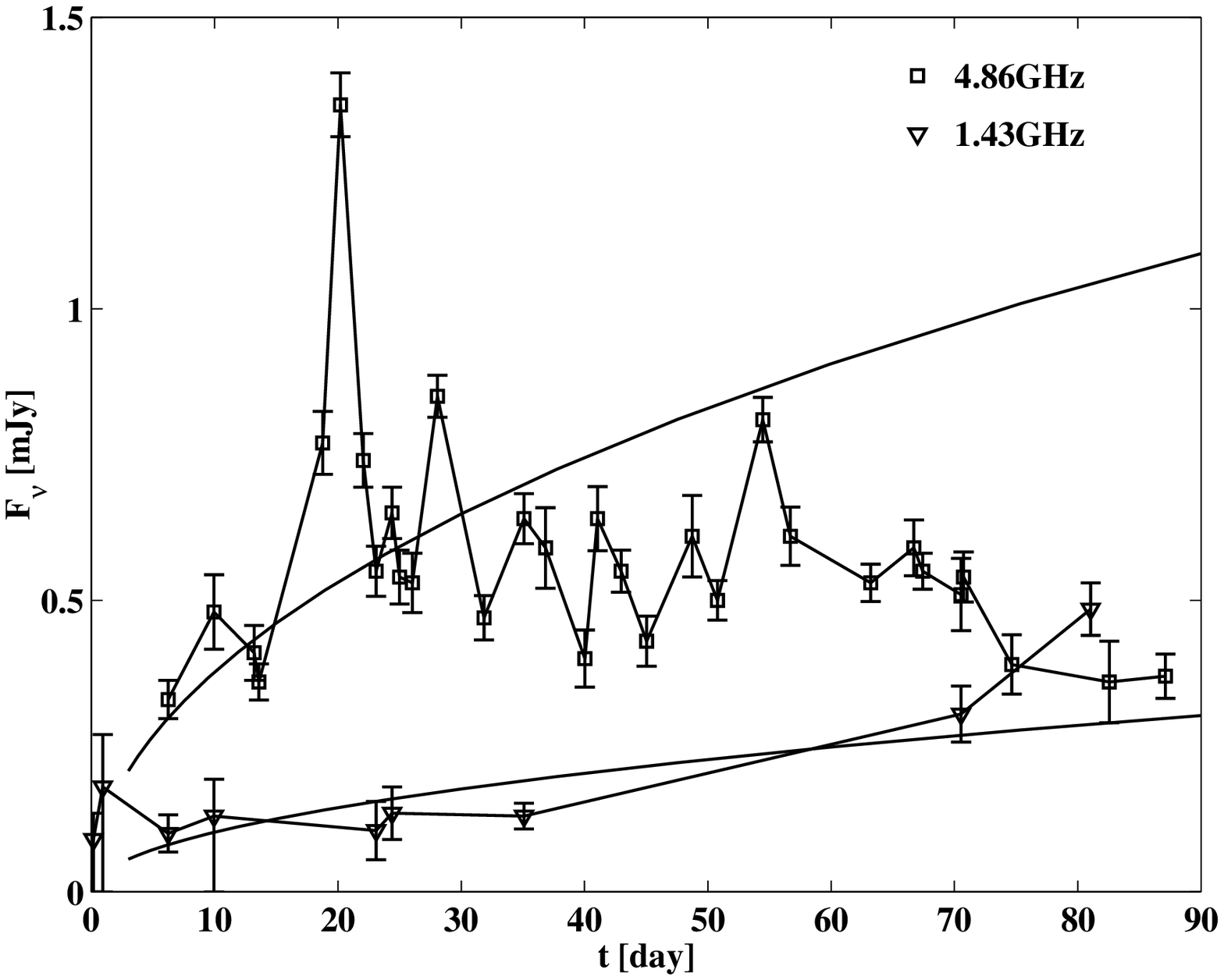}
\figurenum{1a}
\caption{
Light curves of the radio afterglow of GRB970508 at 4.86GHz
and 1.43GHz, compared with the predictions of the adiabatic
fireball model (\cite{AGWb}).
}
\label{fig1a}
\end{figure}

\begin{figure}
\plotone{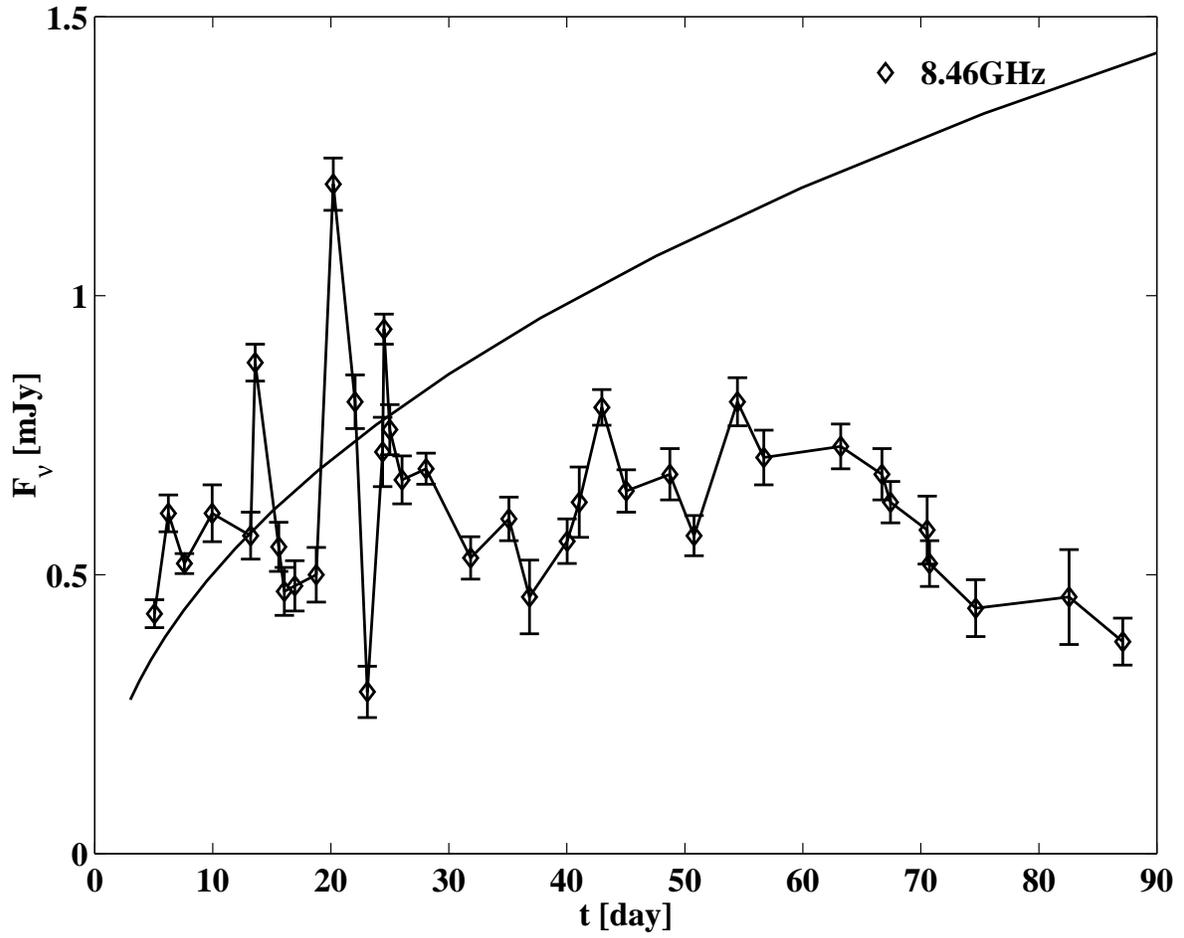}
\figurenum{1b}
\caption{
Light curve of the radio afterglow of GRB970508 at 8.46GHz, compared with 
fireball model predictions (\cite{AGWb}).
}
\label{fig1b}
\end{figure}

\begin{figure}
\plotone{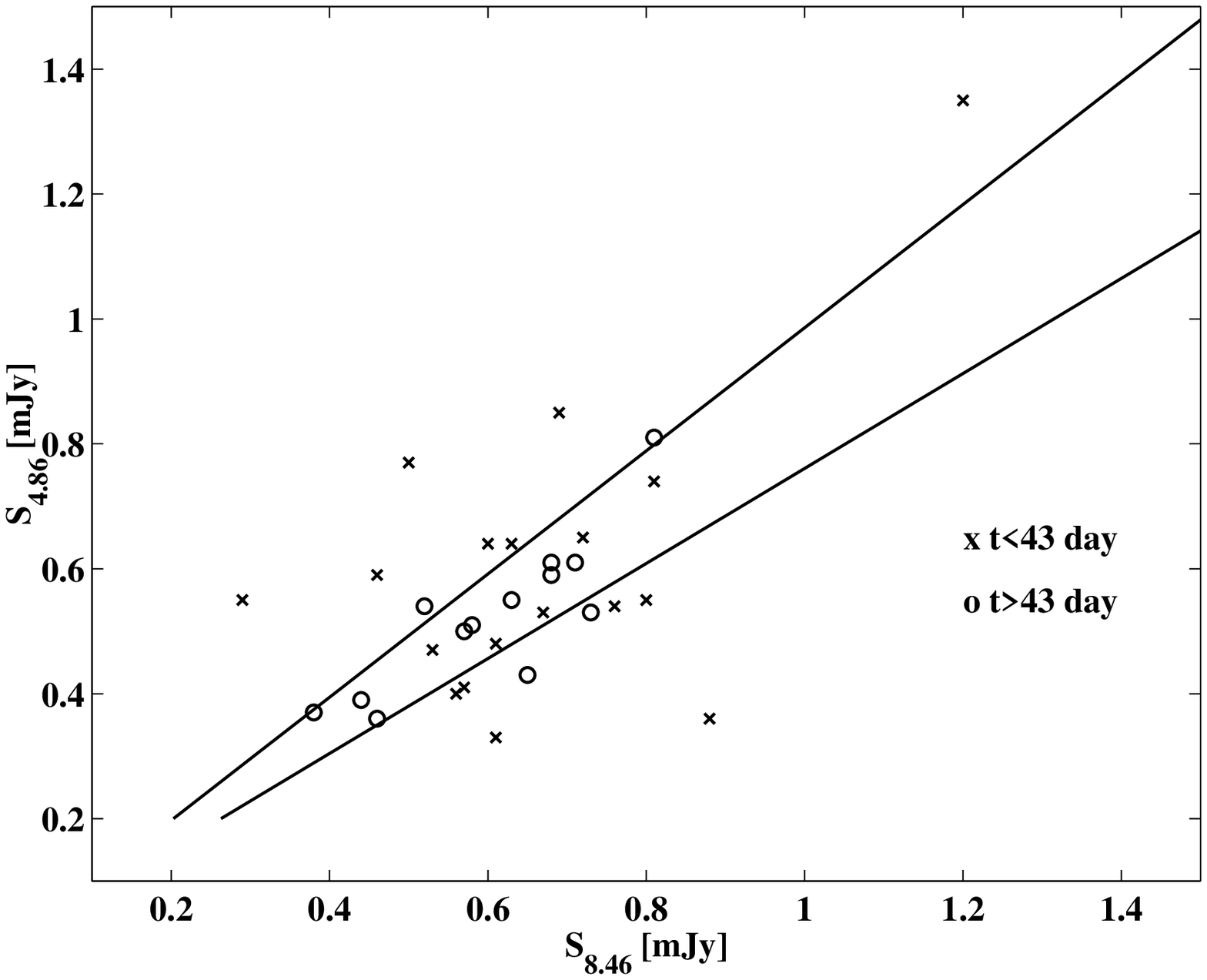}
\figurenum{2}
\caption{
The flux at 4.86GHz as a function of the flux at 8.46GHz,
for $t>43$day and $t<43$day (first/second half of observing time). 
The average value of the flux ratio 
$f\equiv S_{8.46}/S_{4.86}$ for $t>43$day is $\bar{f}=1.17$ and the
standard deviation in $f$ is $\sigma=0.15$ (the region of 1 standard
deviation is bordered by the solid lines). The scatter is consistent 
with that expected due to errors in flux measurement ($\chi^2=17.6$ for the
hypothesis $f=\bar{f}$ for the 13 data points), and $\bar{f}$ is consistent 
with the value expected in the fireball model, $f=(8.46/4.86)^{1/3}=1.20$. 
At delays $t<43$day the average flux ratio is similar, $\bar{f}=1.18$. The
scatter is, however, large, $\sigma=0.45$, and inconsistent with that expected 
from measurement errors ($\chi^2=267$ for the hypothesis $f=\bar{f}$ for the 
18 data points). 
}
\label{fig2}
\end{figure}

\end{document}